\documentclass{article}

\usepackage{arxiv}

\usepackage[utf8]{inputenc}
\usepackage[T1]{fontenc}
\usepackage[authoryear,round]{natbib}
\usepackage{hyperref}
\usepackage{url}
\usepackage{booktabs}
\usepackage{amsmath,amssymb}
\usepackage{graphicx}
\usepackage{microtype}
\IfFileExists{doi.sty}{\usepackage{doi}}{}
\IfFileExists{orcidlink.sty}{\usepackage{orcidlink}}{\providecommand{\orcidlink}[1]{\href{https://orcid.org/#1}{\texttt{orcid.org/#1}}}}

\title{Trust by Context, Not by Design? A Quantitative Study of Data Donation Willingness for Open-Source Civic AI in Switzerland}

\usepackage{authblk}

\author[1]{Sabine Wildemann\thanks{\texttt{sabine.wildemann@aiLights.org}}}
\author[2]{Daniel Ambach\,\orcidlink{0009-0007-7932-5244}\thanks{Corresponding author: \texttt{daniel.ambach@dbuas.de}}}
\affil[1]{aiLights Association, Zurich, Switzerland}
\affil[2]{DBU University of Applied Sciences, Berlin, Germany}

\renewcommand{\shorttitle}{Trust by Context, Not by Design?}
\renewcommand{\undertitle}{Extended preprint version}

\hypersetup{
pdftitle={Trust by Context, Not by Design? Data Donation Willingness for Open-Source Civic AI in Switzerland (extended version)},
pdfauthor={Sabine Wildemann, Daniel Ambach},
pdfkeywords={Data donation, Anonymization, Dynamic consent, AI training data, Privacy calculus, Civic AI},
}

\begin{document}
\maketitle

\begin{abstract}
Civic AI systems are increasingly used to support democratic participation and need to reflect how residents think, ask, and decide. Interactions with such systems may reveal sensitive political views, creating tension between improving AI models and residents' expectations of privacy, consent, and democratic legitimacy. To investigate how this tension influences data donation decisions, a Swiss ballot information chatbot was developed and embedded in an online experiment in which residents asked questions about upcoming ballot initiatives. This study examines the conditions of transparency and user control under which Swiss residents are willing to donate their anonymized chatbot conversations to train an open-source AI model.

A $2\times2$ between-subjects factorial experiment manipulated the presence of a Data Nutrition Label and a granular consent dashboard prior to a donation decision. A custom-built chatbot based on the Apertus-70B large language model was implemented in a multilingual online survey. After interacting with the system, the survey participants decided whether to donate their data. The main outcome variable was the binary donation decision, followed by measures of risk perception, trust, and qualitative decision factors.

The results indicate that neither transparency nor user control significantly affected the data donation decision. Across 205 participants, donation rates were uniformly high across all the condition groups, with an overall rate of 91.7\%. This produced a ceiling effect that left insufficient variance to detect treatment differences. Bayesian robustness checks provided strong evidence of the absence of treatment effects. Although the granular dashboard successfully increased perceived control, it did not lead to higher donation rates. Notably, participants in the high-control condition actively restricted their data-use settings. This indicates that high donation rates coexist with clear boundaries on data use. Qualitative analysis of 120 open-ended responses revealed that participants viewed data donation as a prosocial act motivated by the desire to improve AI tools, support democracy, or contribute to research.

These findings indicate that in non-commercial settings characterized by high institutional trust, perceived public benefit has a stronger influence on data donation decisions than interface design does. Offering residents more control was not an incentive to donate but a way to let them define the terms of their contribution.
\end{abstract}

\keywords{Data donation \and Anonymization \and Dynamic consent \and AI training data \and Privacy calculus \and Civic AI}

\medskip
\noindent\textbf{JEL Classification:} C93, C25, D83

\medskip
\noindent\emph{Note.} This is the extended preprint version. A condensed, journal-length version has been submitted to \emph{AStA Wirtschafts- und Sozialstatistisches Archiv}. Technical implementation details, the full survey instrument, and additional appendix material are available from the corresponding author on reasonable request.

\section{Introduction}\label{sec:intro}

\subsection{Background and context}\label{subsec:background}

As Artificial Intelligence (AI) is now an integral component of our lives, it also changes the relationship between governments and their citizens. We have moved past the era of digitizing paper-based bureaucracy and now integrate systems known as Civic AI. These systems form a digital layer in which municipal decisions about social services, infrastructure, and public safety more and more rely on predictive machine learning models. Civic AI as a concept emerged from discussions about the governance of smart cities \citep{madaio2020}, but nowadays, we discuss how these algorithmic systems can mediate democratic life. If Civic AI is to be more than just an administrative tool, residents must be involved. For Civic AI to serve as a platform for meaningful participation, civic empowerment must occur along at least three dimensions: institutional transparency is required to enable public debate, functional empowerment where AI supports residents in daily tasks, and democratic empowerment to actively involve residents in system design \citep{drobotowicz2023,murrayrust2025}. Taken together, real participation is about giving residents the option to question and shape the algorithms that affect their lives.

The OECD report ``Governing with Artificial Intelligence'' \citep{oecd2025} examined 200 government AI cases and found that civic participation is among the areas where adoption can mature the fastest. One of the results was that most projects never get past the pilot stage because concrete guidance on implementation is still lacking. The report also points out that when people do not understand how these systems work, it erodes accountability and weakens public trust in the government.

Public services can now integrate Large Language Models (LLMs) that have the capability to help residents to decode complicated legal papers and assist them in finding relevant details about voting procedures, healthcare access, or tax-related information. They also enable equal access to political information through their technological features. This is especially relevant for countries like Switzerland, which are built on a tradition of direct democracy and a linguistically diverse landscape. Swiss citizens participate in ballot initiatives several times a year and need to navigate voting information across four national languages. An LLM-based tool can be very helpful in overcoming these barriers. But these systems should not be developed in isolation from the citizens they are meant to serve. While foundation models learn factual knowledge during pre-training, they can be fine-tuned to be useful by learning from real citizen dialogues \citep{ouyang2022}. These original data can help researchers develop effective models that address the concerns and needs of residents.

\citet{sieber2025} reviewed over 3{,}000 articles that investigated practices in public involvement. Their findings show that most studies treat public participation either as a consequence of government automation, as a process shaped by AI technology, or as a technocratic means of building trust. They also found that current research on how public involvement affects political power dynamics or citizen influence in decision-making processes remains limited. Sieber et al. make the point that public participation needs to be meaningful, it should let citizens exercise control over AI systems. This kind of creative participation, where residents do not just use systems but help to shape them to serve civic needs, is known as data donation.

Following the idea of Participatory Design (PD), citizens are no longer passive users of AI systems. The concept has been recognized as the most prevalent form of public participation in AI policy-making \citep{delgado2021}. Residents take an active role when donating their data and support shaping how the system understands and responds to their actual needs.

For open-source projects like the Swiss AI project Apertus, developed by ETH Zurich and EPFL \citep{hernandezcano2025}, the idea of civic contribution could be extended from data collection to co-production. People could actively shape what the AI optimizes for. By donating their data, they would help define what ``good'' means for AI. \citet{lee2019} described this with an example where community members of a food-delivery nonprofit collectively defined what the matching algorithm should prioritize. For Civic AI, such co-production could create a self-reinforcing cycle. If enough people contribute, the model improves over time, thereby enabling broader political participation. This builds the kind of trust that keeps the cycle going.

This study applies the human-centered AI framework from \citet{shneiderman2020}. It bases the trustworthiness of AI systems on two elements, user control and transparency. To make the Participatory Design process truly meaningful, the interface design must support these two elements. However, what are the mechanisms that achieve this effect?

\subsection{Problem statement}\label{subsec:problem}

The development of Civic AI systems faces a fundamental paradox. Large-scale foundation models require enormous amounts of high-quality, context-specific data to remain accurate and culturally cognizant. As \citet{ouyang2022} showed, these systems only become useful when fine-tuned on real user interactions. Through this process, models learn to understand what users ask for and how to respond in a way that is appropriate. For Swiss Civic AI, this means fine-tuning models on what residents actually ask when engaging in direct democracy: What are the uncertainties of people weighing the pros and cons of ballot initiatives? What are the regional quirks of civic concern? How are Swiss German speech patterns relevant to the way questions are asked? Relying on synthetic queries falls short because they cannot mimic these nuances and behaviors, and the model might optimize for imagined needs over real ones, generating technically correct but useless answers. However, recent data-centric breakthroughs, such as Microsoft's Phi-4 \citep{abdin2024}, demonstrate that high-performance reasoning can be achieved by using human data as ``seeds'' for high-density synthetic reasoning pipelines. This suggests a technical pathway where the social license for data donation is used not for volume but for cultural anchoring.

While this seeding approach reduces volume, the inputs required, comprising inquiries about political initiatives and voting intentions, remain sensitive personal information under Swiss law. The revised Federal Act on Data Protection \citep{fadp2023}, aligned with the European Union's General Data Protection Regulation \citep{gdpr2016}, states that queries revealing ``political views or activities'' require explicit consent and robust governance safeguards. The technical necessity for training data is in direct conflict with legal and societal expectations regarding privacy, consent, and user autonomy.

To address the friction between technical requirements and individual rights, it is clear that the current data consent process does not bridge the gap. Clickwrap agreements, which reduce the consent of an individual to a single click on an ``I agree'' button, are the standard practice across most online services. They typically protect companies from problems. Many people do not always know what the agreement says or what they agree to when they accept it. This problem worsens because of what \citet{acquisti2015} describe as the Transparency Paradox, which is discussed in more detail in Section~\ref{subsec:heuristic}. When privacy policies list all the small technical rules to comply with the law, users find it hard to know their rights. These privacy policies often confuse individuals and do not help them much, and attempts at full transparency paradoxically leave people less informed and feeling that they have less control over their data.

However, even when people say they care about privacy, they often give away their data without hesitation. This is a well-documented pattern known as the Privacy Paradox \citep{dinev2006}. This pattern has mostly been observed in commercial settings, where individuals trade personal information for free services. In civic settings that refer to political views, people are likely to approach data sharing more cautiously. This raises the question of which consent mechanisms can address this issue.

Beyond these challenges, it is necessary to discuss what democratic participation in AI means. The current ethical standards for AI systems have generally conceived democratic participation as protective or ex-post auditing, which examines the decisions generated by algorithms \citep{jobin2019}. In contrast, democratic participation can also function as an ex-ante creative process that enables residents to provide data that help AI systems develop civic knowledge \citep{baum2001}. Data donation for open-source systems like Apertus could be considered as this form of creative democratic participation, where the resident is part of building the democratic knowledge of the AI system, as opposed to being subjected to it. But we need to consider that the ability to participate is not the same for everyone. \citet{baum2001} identifies participation in public decision-making as a basic element of democratic citizenship but also notes that designing inclusive processes is a challenge. In particular, low-income citizens face difficulties in participation. If a Civic AI interface is complex or difficult to understand, it will mostly attract well-educated users, and the resulting training data will reflect only their perspectives.

A sovereign architecture built on Apertus provides a necessary foundation, but it does not resolve this dilemma on its own. Technical transparency, while important, is not sufficient; it needs to be translated into what has been called a social license to operate. Originally developed in the context of the mining industry, this concept describes the level of acceptance and approval that stakeholders grant to an organization or project, based on the cumulative foundations of legitimacy, credibility, and trust \citep{thomson2011}. \citet{leonard2018} extended this concept to the digital domain and argued that data-driven applications and AI systems face similar challenges. They require not only legal compliance but also a broader social trust that communities are willing to grant or withhold. In the context of this study, a social license means the collective willingness of Swiss residents to contribute to a shared digital infrastructure, not because they are legally required to, but because they find the project to be legitimate and trustworthy. To achieve this, design mechanisms that reflect Swiss values of civic competence, personal voice, and individual autonomy are required.

\subsection{Research question and hypotheses}\label{subsec:rq}

The following research question guides this study, along with three hypotheses derived from the theoretical framework presented in Section~\ref{sec:theory}:

\medskip
\noindent\textit{Under what conditions of transparency and user control are Swiss residents willing to donate their anonymized chatbot queries to train a Swiss open-source AI model?}
\medskip

\noindent\textbf{H1 (Transparency Hypothesis).} The presence of a standardized Data Nutrition Label (visualizing provenance and safeguards) significantly increases the probability of data donation compared to a standard text-based disclosure.

\noindent\textbf{H2 (Control Hypothesis).} The provision of a granular consent dashboard (allowing restrictions on scope and retention) significantly increases the probability of data donation compared with a binary (accept/decline) choice.

\noindent\textbf{H3 (Interaction Hypothesis).} There is a positive interaction effect between perceived transparency and perceived user control, such that the highest donation rates occur when both mechanisms are present simultaneously.

These hypotheses were tested through a $2\times2$ between-subjects factorial experiment that integrated a functional Swiss ballot chatbot powered by the Apertus-70B model \citep{hernandezcano2025}.

\section{Theoretical background}\label{sec:theory}

When residents are asked to decide whether to share their data, they consider the expected advantages and risks. This is not a fully rational process. People have limited attention and usually do not read the privacy policies. They are highly influenced by how choices are presented and rely on shortcuts and intuition. The interface as well has a big impact. \citet{kehr2015} termed this a bounded privacy calculus. They argue that the situation-specific assessment of risks and benefits is constrained by pre-existing dispositions and the limited cognitive resources of an individual. In the following, three behavioral theories are described to ground the hypotheses: Privacy Calculus, Heuristic Information Processing, and Psychological Ownership through Dynamic Consent.

\subsection{Privacy Calculus and the Privacy Paradox}\label{subsec:calculus}

The Privacy Calculus model \citep{culnan1999,dinev2006} posits that individuals make rational trade-offs when deciding whether to disclose personal information. People weigh the perceived benefits of disclosure (utility, personalization, and social connection) against the perceived risks (surveillance, misuse, and loss of control). They share their data when they feel the benefits are worth it and hold back when the risks feel too high.

The trade-off is well understood in commercial contexts. Individuals share their personal data in exchange for convenience, better recommendations, and access to free services and products. However, the civic context operates under different conditions. The reward for donating data to a public AI model is not personal convenience but the improvement of a public good that serves a broader community. Research on data donation for medical research suggests that prosocial motivations can drive data sharing, but this willingness is highly contingent on trust in the data recipient \citep{skatova2019}. When the recipient is a public institution or academic research project rather than a commercial entity, donation rates tend to increase, but only when people feel that the public actually benefits from it and that the institution handling their data can be trusted to do so.

The Privacy Paradox complicates this rational model. Empirical studies show that individuals strongly worry about privacy, but they still share their personal details when using services \citep{norberg2007}. Several explanations have been proposed for this. Individuals tend to apply temporal discounting, opting for immediate rewards rather than considering potential future risks \citep{acquisti2005}. In addition, individuals struggle to process complex information and fall back on mental shortcuts that may not reflect what they really want. In commercial settings, companies take advantage of this and design consent interfaces with dark patterns or pre-set defaults and tend to flood users with information rather than help them make informed choices.

As soon as sensitive political data are involved, the Privacy Paradox may not work the same way. \citet{hoffmann2023} found that in political participation contexts, the link between privacy concerns and actual behavior is neither consistently negative nor positive. This depends heavily on the context. Political data carry risks that go beyond those of commercial data. There is a history of surveillance and persecution based on political beliefs, and more recently, the threat of microtargeting in elections.

This study therefore asks if transparency and control can encourage people to donate their data, even in a context where political sensitivity may make them more cautious than the Privacy Paradox would predict.

\subsection{Heuristic information processing and the Transparency Paradox}\label{subsec:heuristic}

The Transparency Hypothesis (H1) builds on \citet{chaiken1980} Heuristic-Systematic Processing Theory. This model describes two ways in which people process information. Sometimes, people go through everything carefully. They read the privacy policy line by line, weigh each clause, and then decide. However, most of the time, people take shortcuts. They look at visual cues, judge whether the source seems credible, and apply simple rules of thumb to make quick decisions without much mental effort.

Research shows that people do not carefully read privacy information. Privacy policies are rarely read in their entirety. \citet{mcdonald2008} estimated that reading every privacy policy a person encounters in a year would take approximately 76 working days. Instead, people scan for signals such as logos, checkmarks, and color codes. Anything that appears trustworthy serves as a stand-in for actual understanding. This dynamic reflects the Transparency Paradox \citep{acquisti2015}: adding more details to a privacy policy does not help people understand it better. It overwhelms them, and they stop engaging altogether. The legal requirement for comprehensive disclosure ultimately works against its purpose.

A Data Nutrition Label (DNL) can be used to navigate this paradox and apply what is known as heuristic transparency. Developed by the MIT Data Nutrition Project \citep{holland2018,chmielinski2022} and extended to AI models by \citet{stoyanovich2019}, the DNL adapts a familiar food nutrition label format to show key dataset and model characteristics. Through elements such as visual chunking, iconography, and color-coding, key characteristics are easy to grasp. Section~\ref{subsec:stimulus} describes the custom-built DNL used in this study. H1 tests if these visual cues reduce perceived risk enough to increase donation rates compared to standard text disclosure.

\subsection{Psychological Ownership and Dynamic Consent}\label{subsec:ownership}

The Control Hypothesis (H2) draws on the Psychological Ownership theory \citep{pierce2003,peck2009}. According to this theory, individuals develop a sense of ownership over things, even intangible ones, such as personal data. This can occur when they control something, invest effort in it, or become closely acquainted with it. When individuals feel ownership of their data, they become more protective of it. However, it also makes them more open to sharing, provided the conditions feel fair and they have a say in how it happens.

Configuring something creates a sense of ownership. Individuals invest cognitive effort when adjusting settings or defining boundaries in an interface. Such efforts lead to attachment. \citet{pierce2003} showed that this kind of control and personalization increases satisfaction and engagement because individuals have shaped the outcome themselves. In the context of data donation, this is important and makes their role active. Rather than having their data taken from them, they decide what happens to it. This transforms the relationship from one of extraction to one of partnership and it fundamentally changes how the Privacy Calculus plays out.

Dynamic Consent provides an operational framework that addresses this problem. \citet{kaye2015} originally developed it for biomedical research. Instead of asking for consent once and covering everything in a single agreement, it creates an ongoing process. Individuals decide what types of data they share and for what purposes. They can adjust these preferences over time, see how their data are being used, and withdraw consent at any point. With this approach, two main problems in conventional clickwrap consent systems are addressed: data often end up being used for purposes that nobody anticipated when consent was given, and individuals rarely get a meaningful alternative. It is usually to accept everything or get nothing.

\citet{andreotta2022} argue that dynamic consent is especially relevant for AI, where data may be reused across multiple training cycles. The granular consent dashboard in this study puts dynamic consent into practice by allowing participants to configure how their data are used. H2 tests whether this configurability increases donation rates compared to a binary choice by increasing perceived agency and Psychological Ownership.

\subsection{The Interaction of Transparency and Control}\label{subsec:interaction}

H3 proposes that transparency and control are synergistic, which means that their combination produces effects greater than the sum of their parts. Each mechanism targets a different barrier and is interdependent on the other to achieve its full effect.

Transparency decreases the information gap between what a system operator knows and what a user understands. Individuals need clear information to assess risks properly because otherwise they must either trust blindly or assume the worst. The absence of control in transparent systems creates a new problem, which researchers call informed helplessness. Individuals who understand how their data will be used but cannot influence the terms may experience resignation rather than trust \citep{shklovski2014}. The practice of storing political inquiries on foreign servers without user access to change this practice does not build trust and may actually reduce their willingness to engage.

Conversely, control without transparency triggers the Control Paradox \citep{brandimarte2013}. When individuals are offered configuration options they do not understand, the cognitive load of decision-making increases anxiety rather than agency. Choosing between ``Swiss servers'' and ``EU servers'' is meaningless if the individual does not understand why server location matters for data protection.

The combination of heuristic transparency and granular control resolves these two problems. The Data Nutrition Label provides cognitive scaffolding that makes control options meaningful. Participants who have scanned the label understand that ``Swiss servers'' relates to data sovereignty and legal jurisdiction; they can make the choice with confidence. Transparency enables informed control, and control can make transparency actionable. \citet{murrayrust2025} formalized this synergy for Civic AI, arguing that meaningful transparency must be not only technically producible but also connected to lived experiences and actionable for the people. Without actionability, transparency produces what they term informed helplessness. Without comprehension, control creates anxiety rather than agency. This aligns with \citet{shneiderman2020} Human-Centered AI framework.

The theoretical synergy between transparency and control can be operationalized through auditable provenance, a concept exemplified by the MixtureVitae corpus \citep{nguyen2025}. By categorizing data into risk-based licensing tiers, Civic AI systems move beyond static labels towards verifiable, permissive-first transparency. This technical approach ensures that user control options are grounded in auditable data practices, thereby meeting residents' privacy expectations.

Consequently, H3 predicts that the highest donation rates will occur in condition D (DNL + dashboard), where users receive transparency to reduce risk perception and control to increase their perceived agency.

\section{Research design and methodology}\label{sec:methods}

The following section describes the study design, including the experimental conditions and the stimuli. It explains the technical implementation and sovereign system architecture and how the sample and data collection were managed. The section concludes with a description of the measurements and analytics.

\subsection{Study design}\label{subsec:design}

\begin{figure}[ht]
\centering
\includegraphics[width=0.95\textwidth]{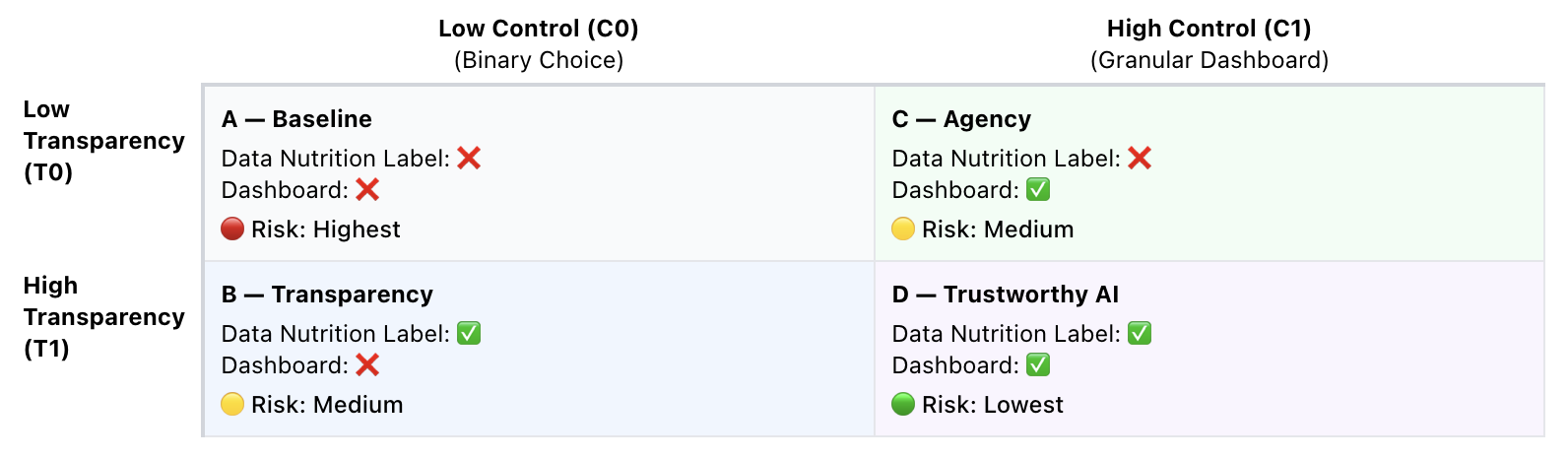}
\caption{The $2\times2$ factorial design with stimulus assignment and the resulting risk pattern across the four conditions}\label{fig:design}
\end{figure}

A $2\times2$ between-subjects factorial experiment was selected to test the causal effects of transparency and user control on data donation behavior. The two independent variables, \emph{Transparency} (low vs.\ high) and \emph{Control} (low vs.\ high), were fully crossed, yielding four experimental conditions (Figure~\ref{fig:design}). Participants were randomly assigned to one condition using block randomization, which ensured balanced cell sizes.

The experiment was embedded in a functioning civic AI application. The participants interacted with a chatbot powered by the Swiss Apertus-70B model. The chatbot provides information on Swiss ballot initiatives using data from the Swissvotes database \citep{swissvotes2025}. This contextualized approach increases ecological validity because participants make donation decisions after interacting with the system and not based on abstract descriptions of a system or a survey.

\begin{figure}[ht]
\centering
\includegraphics[width=0.95\textwidth]{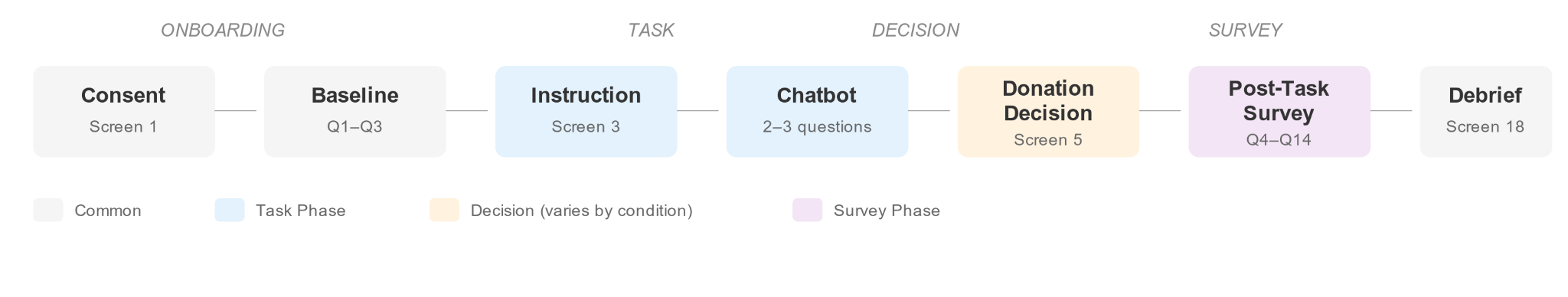}
\caption{The nine-phase study protocol, from eligibility screening and informed consent through the chat interaction and the condition-specific donation decision to the post-task survey and debriefing}\label{fig:protocol}
\end{figure}

This study followed a nine-phase protocol: (1)~landing page with introduction and eligibility screening (18+, Swiss resident); (2)~informed consent with participation confirmation; (3)~baseline measures (technology comfort, privacy concern, ballot familiarity); (4)~chatbot instruction with task explanation and example questions; (5)~chat interaction in which participants ask two to three questions about ballot initiatives; (6)~the condition-specific donation decision; (7)~confirmation of the decision; (8)~a post-task survey with manipulation checks, risk perception, and trust measures; and (9)~debriefing that disclosed the simulated donation and the study purpose (Figure~\ref{fig:protocol}).

In the chatbot interaction phase (Phase~5), participants were given access to an AI-based chatbot before making their donation decisions. To continue with the survey, participants had to interact with the chatbot at least twice. This allowed them to understand how the system works and evaluate whether its responses were useful. A critical design decision concerns the measurement of the dependent variable. This study measured willingness to donate rather than actual data donation, following established approaches in behavioral privacy research \citep{norberg2007}. When participants chose to donate their data, their chat conversations were not stored or used for training purposes. The donation process was completely simulated. This design choice serves several purposes. It resolves ethical concerns regarding the collection of sensitive political data under potentially incomplete informed consent. It also allows for clear debriefing without deception regarding data handling. And it focuses the research question on the psychological mechanisms driving donation decisions rather than on post-hoc behavior that may be influenced by extraneous factors. In Phase~9, after the participants completed the survey, they were fully debriefed and informed that the donation was simulated and that the chat data were not stored. This approach aligns with the FADP requirements for transparent data processing and allows for causal inference regarding how interface design affects user behavior.

\subsection{Technical implementation and system architecture}\label{subsec:tech}

The system consists of a React frontend, a Node.js backend that handles both chat orchestration and data collection, a Python service for ballot data enrichment, and Apertus as an external LLM API.

\paragraph{Sovereign Swiss infrastructure and hosting.} A core approach to implementation was the elimination of dependency on foreign AI providers. The technical stack is hosted on Infomaniak's Jelastic Cloud infrastructure within the Swiss jurisdiction. By hosting all components, from the frontend to the database, in Switzerland, the study ensured that session metadata remained protected by the revised Federal Act on Data Protection (FADP). The chatbot runs on the Apertus-70B model, a Swiss large language model for natural language processing (NLP), hosted through Infomaniak's AI API. This setup guarantees that participants' queries are processed locally and are not subject to the terms of service of international and commercial providers.

\begin{figure}[ht]
\centering
\includegraphics[width=0.9\textwidth]{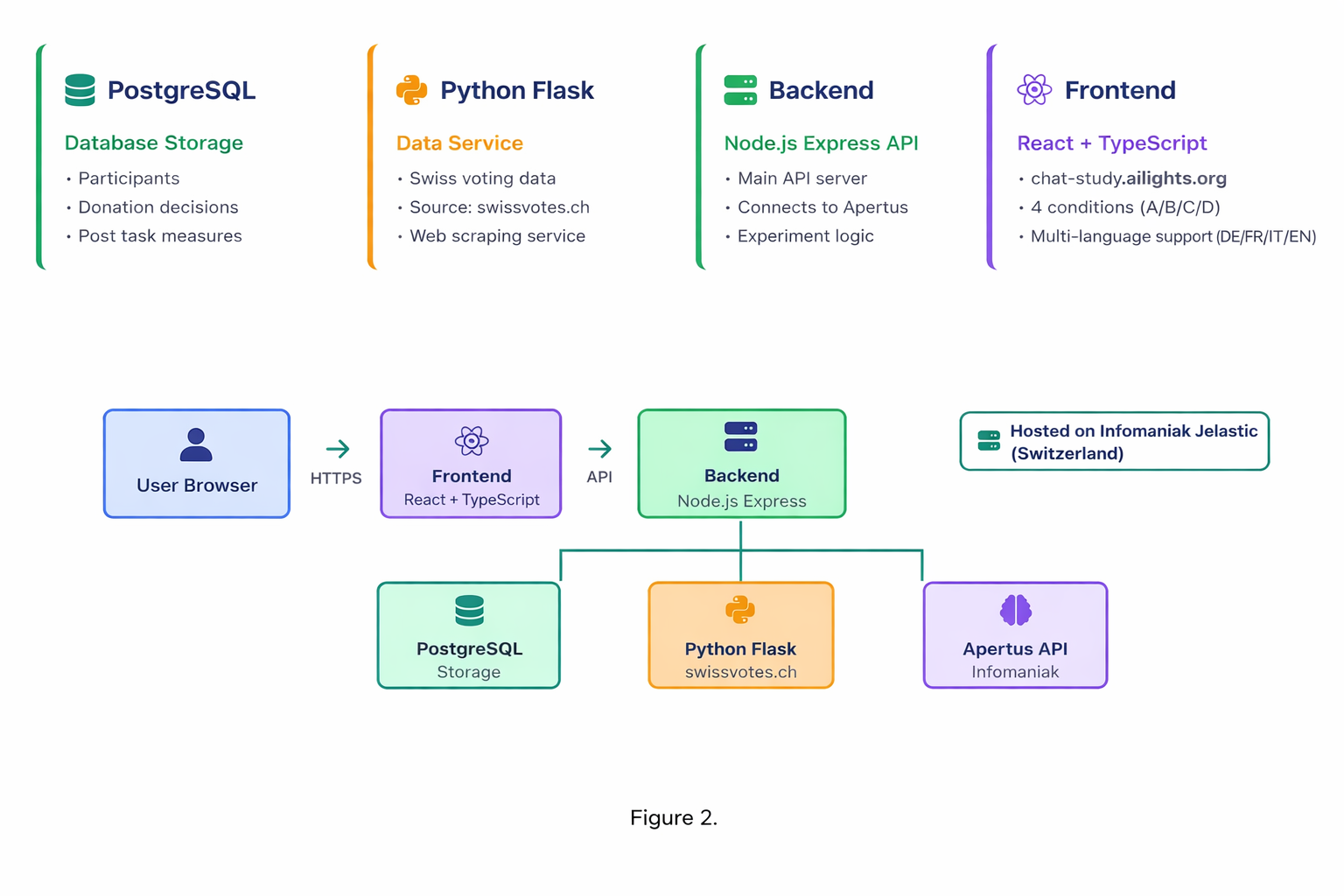}
\caption{Overview of the system architecture and the data flow between the frontend, backend, ballot-data service, and the Apertus API, hosted on sovereign Swiss infrastructure}\label{fig:arch}
\end{figure}

\paragraph{Information pipeline and LLM grounding.} The backend was developed using Node.js and Express and managed the interaction between the user and the AI stimulus through a multi-layered pipeline. A Python Flask microservice implements contextual grounding by directly injecting verified external data into the system prompt. This approach combines the built-in knowledge of the LLM with structured civic data to avoid hallucinations, improve factual accuracy, and makes it possible to trace the model's outputs \citep{gao2023}. The service fetches official data from the Swissvotes database to ground the Apertus-model responses; the complete list of API endpoints is available on request. When a participant submits a query, the system looks up the upcoming votes from the cached Swissvotes dataset and appends the relevant ballot context to the system prompt. The system has an optional feature that downloads official ballot PDF files (for example, \emph{Abstimmungstext}, \emph{Botschaft des Bundesrats}) before extracting the data for system prompt enrichment. The ballot context includes vote dates, recommendations from parties, and Federal Council positions and is injected directly into the system prompt. To keep response times low without sacrificing data freshness, a two-tier caching strategy was implemented: the Swissvotes CSV dataset was locally cached for seven days, and the extracted text from the ballot PDFs was permanently cached in a dedicated directory. The Apertus-70B model was configured with a temperature of 0.1 and nucleus sampling (top\nobreakdash-$p = 0.9$) to ensure factual and deterministic responses, which is very important in the context of public services. The React-based frontend uses react-i18next to offer the interface in German, French, Italian, and English. The system includes a state-managed user journey that only triggers the donation request after a minimum of two successful chat interactions, ensuring that participants have a firsthand experience with the system's utility. Figure~\ref{fig:arch} provides an overview of the system architecture and data flow between the components.

\paragraph{Data infrastructure and privacy-by-design.} The data infrastructure implements privacy-by-design principles through data minimization, purpose limitation, and architectural separation. \emph{Data minimization:} the system only collected the data needed to test the three hypotheses. Survey responses, manipulation checks, donation decisions, and dashboard configurations were stored in a PostgreSQL database hosted in Switzerland. None of the chat conversations of the survey participants were saved to disk, and messages passed through the backend to the Apertus API without being written to disk or any persistent store. This architectural choice guaranteed that even when participants chose to donate their data, their actual queries were not stored. \emph{Separation of concerns:} the architecture isolates the experimental logic from AI inference. The Node.js backend manages the session state, condition assignment, and survey data collection, whereas the Python Flask service handles only public Swissvotes data retrieval. API calls to the Apertus model passed through the backend, where operational metadata (token usage, response times, and model identity) were logged to verify that the provider was connected to the correct model. No message content was recorded in the database. \emph{Anonymization:} participants were identified solely by a randomly generated session UUID. No personally identifiable information (name, email, or IP address) was collected or stored. Browser fingerprinting is used exclusively for duplicate detection and is not linked to the survey responses. The resulting dataset contained only condition assignment, survey responses, and timestamps, none of which could be traced to individual participants. This infrastructure enables FADP-compliant research on sensitive political topics without creating privacy risks.

\subsection{Experimental conditions}\label{subsec:conditions}

The $2\times2$ factorial design yields four experimental conditions that systematically vary in terms of transparency and control mechanisms presented during the donation decision (Phase~6; Figure~\ref{fig:design}). Participants in all conditions completed the same chatbot interaction and received the same donation request; only the accompanying interface elements differed.

\emph{Condition A (baseline)} serves as a condition with low transparency and low control: participants received a brief textual description of how their anonymized data would be used, followed by a binary choice to donate or decline. Condition A mirrors the standard consent interface found in most online services.

\emph{Condition B (Transparency)} introduces the Data Nutrition Label (DNL) while maintaining a binary choice format. The DNL presents factual information about the AI model through a structured visual display, allowing participants to assess data-handling practices using heuristic cues rather than lengthy text. Condition B tests H1 by isolating the effect of increased transparency.

\emph{Condition C (Agency)} provides a granular consent dashboard without the DNL. Participants configure their data-use preferences through the dashboard before making their decisions (see Section~\ref{subsec:stimulus}). This condition tests H2 by isolating the effect of user control and maintaining transparency at the baseline.

\emph{Condition D (Trustworthy AI)} combines both mechanisms, presenting the DNL followed by the dashboard. It tests H3 by examining whether combining transparency and control produces a stronger effect than either one alone.

The condition labels reflect their theoretical function: ``Baseline'' establishes reference rates; ``Transparency'' and ``Agency'' isolate each main effect; ``Trustworthy AI'' operationalizes \citet{shneiderman2020} framework, where both pillars are present. In conditions with the dashboard, the response buttons remained disabled until all configuration questions were answered.

\subsection{Stimulus design}\label{subsec:stimulus}

Two stimuli were developed to manipulate transparency and control: a Data Nutrition Label (DNL) and a granular consent dashboard.

\begin{figure}[ht]
\centering
\includegraphics[width=0.55\textwidth]{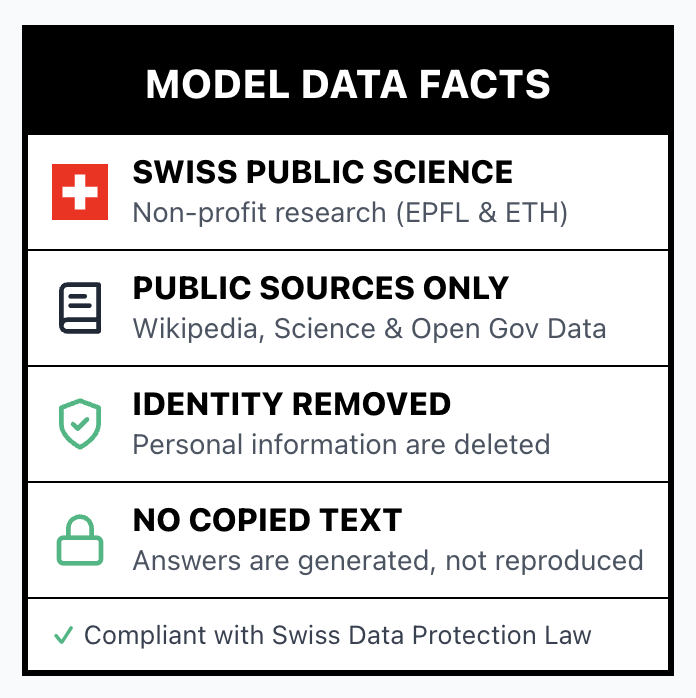}
\caption{The Data Nutrition Label displayed in conditions B and D}\label{fig:dnl}
\end{figure}

\paragraph{Data Nutrition Label.} Though the DNL follows the conceptual framework established by \citet{holland2018}, the visual artifact and categorical taxonomy used in this study were independently designed to align with the specific requirements of the Swiss civic context. The implementation utilizes visual chunking and iconography to communicate four key model attributes: institutional provenance, data sources, privacy protection, and output generation (Figure~\ref{fig:dnl}). At the bottom, the label indicates compliance with the FADP.

\begin{figure}[ht]
\centering
\includegraphics[width=0.6\textwidth]{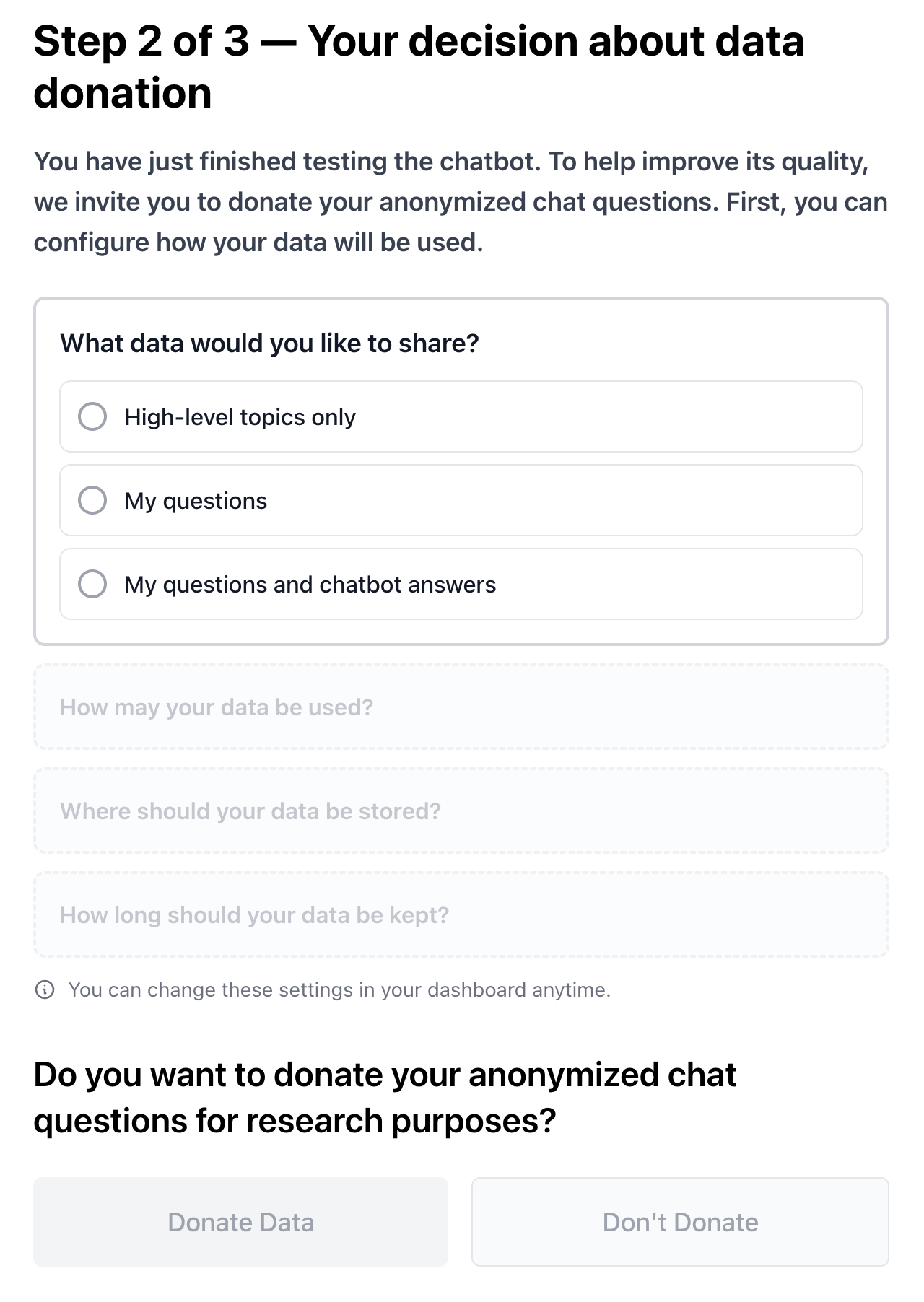}
\caption{The granular consent dashboard displayed in conditions C and D}\label{fig:dashboard}
\end{figure}

\paragraph{Granular consent dashboard.} The dashboard operationalized dynamic consent \citep{kaye2015} through four configuration questions (Figure~\ref{fig:dashboard}): (1)~the data scope determined the granularity of the shared content: high-level topics only, the participant's questions, or full conversations, including chatbot answers; (2)~permitted purposes distinguished between research and commercial applications; (3)~storage locations offered Swiss-only servers, Swiss or EU servers, and no preference; and (4)~retention periods included deletion once the research purpose was fulfilled, six months, one year, or indefinite storage. Each question presented graduated options. To reduce cognitive overload, the questions were answered sequentially. All four questions were visible, but only the current question was active, and subsequent questions remained grayed out until the previous question was answered. The response buttons were disabled until all options were selected.

\subsection{Sample, data collection and data preparation}\label{subsec:sample}

\paragraph{Target population.} The target group for the survey were adults aged $\geq 18$ years who live in Switzerland. No restrictions were placed on citizenship, voting eligibility, or language proficiency beyond the self-assessed ability to complete the survey in one of the four available languages (German, French, Italian, and English).

\paragraph{Sample size.} A target of $N = 200$ (50 per condition) was determined using power analysis. With $\alpha = .05$ and an assumed medium effect size ($\mathrm{OR} \ge 2.0$), this provides adequate power ($\approx 0.75$) to detect the main effects in logistic regression. This falls below the conventional threshold of 0.80, indicating that the study accepts a slightly elevated risk of Type~II error.

\paragraph{Recruitment.} Participants were recruited through two channels between 5~January and 5~February 2026. Approximately half of the sample was recruited through convenience sampling using LinkedIn, university networks, and personal contacts. Additional participants were recruited through the Swiss panel provider TestingTime to obtain a more balanced and representative sample. These participants were compensated with CHF~3.00. All other participants took part voluntarily.

\paragraph{Data collection.} Participants accessed the study via a landing page, gave informed consent, completed baseline measures, interacted with the chatbot, made their donation decision, and completed a post-task survey. A single run of the survey took approximately eight minutes. To prevent duplicate participation, a browser fingerprint was checked against previous completions; participants who had completed the study within seven days were blocked. Prior to data collection, the study interface was pre-tested with seven users to verify clarity and functionality on desktop and mobile devices.

\paragraph{Data preparation.} The raw data was stored in a PostgreSQL database containing participant metadata, condition assignments, survey responses, and donation decisions. For the analysis export, personally identifiable information was removed: session UUIDs were replaced with sequential numeric identifiers, email addresses were deleted, and all timestamps were dropped to prevent participant fingerprinting through temporal patterns. The resulting dataset contained 23 variables across 291 records. The survey interface recorded browser locale codes rather than standardized language identifiers. Because participants accessed the study from different browser configurations, the language field contained 14 distinct locale variants (e.g., de-DE, de-CH, gsw, en-GB, en-US, en-AU, fr-CH, es-ES). These were mapped to five standard categories, namely German, French, Italian, English, and other, to enable demographic reporting. Swiss German (gsw) was grouped with German; non-study languages such as Spanish were classified as other. The original locale codes were retained in the dataset for traceability. The donation decision was collected and stored as a text value (donate/decline) and was encoded as a binary indicator (1~=~donate, 0~=~decline). For participants in the dashboard conditions (C and D), the granular consent configuration was stored as a JSON object and parsed into four separate categorical variables: data scope, permitted purpose, storage location, and retention period.

Participants were excluded if they did not complete the post-task survey ($n = 79$) or failed the attention check, which asked them to recall the chatbot topic with ``Swiss ballot initiatives'' as the correct answer ($n = 7$). No records had missing condition assignments or donation decisions. The final analytic sample comprised $N = 205$ (Table~\ref{tab:flow}). All survey items used forced-response formats with predefined option sets, so no item-level missing data or out-of-range values were present in the analytic sample; no imputation was necessary. To assess whether dropout was related to experimental condition, the 79 incomplete cases were examined by condition assignment. A chi-square test indicated no significant association between condition assignment and survey non-completion, $\chi^2(3) = 5.18$, $p = .159$, $V = .13$, suggesting that attrition was unrelated to the experimental manipulation.

\begin{table}[ht]
\centering
\caption{Sample flow and exclusions}\label{tab:flow}
\begin{tabular}{@{}lcc@{}}
\toprule
Stage & $N$ & Excluded \\
\midrule
Initial participants & 291 & 79 \\
After incomplete-survey exclusion & 212 & 7 \\
After failed attention-check exclusion & 205 & -- \\
\textbf{Final analytic sample} & \textbf{205} & -- \\
\bottomrule
\end{tabular}
\end{table}

\subsection{Measurement and analysis}\label{subsec:analysis}

\paragraph{Dependent variable.} The primary outcome was the binary donation decision (1~=~donate, 0~=~decline). This behavioral measure captured whether participants agreed to share their anonymized chatbot conversation for AI training purposes.

\paragraph{Manipulation checks.} Two manipulation checks were used to confirm that the experimental conditions worked as intended. Perceived transparency (MC-T) was measured using two items that asked participants whether they felt informed about how their data would be handled. Perceived control (MC-C) was measured using two items that asked participants whether they felt they could influence how their data would be used. All items were rated on a six-point Likert scale ranging from 1 (strongly disagree) to 6 (strongly agree); the complete baseline and post-task instruments are available from the corresponding author on reasonable request. Composite scores were calculated as the mean of the two items for each construct. Scale reliability for the three two-item composites was assessed using the Spearman--Brown coefficient: MC-T ($\rho = .86$), MC-C ($\rho = .81$), and OUT-RISK ($\rho = .81$), all indicating adequate internal consistency. For the manipulation to be considered successful, participants in the transparency conditions (B, D) should report higher MC-T scores than those without transparency (A, C), and participants in the control conditions (C, D) should report higher MC-C scores than those without control (A, B).

\paragraph{Outcome measures.} Two additional constructs were assessed: perceived risk (OUT-RISK) captured concerns about the potential misuse of donated data through two items, and trust in the research initiative (OUT-TRUST) was measured using a single item. An open-ended question asked participants what mattered most in their donation decisions, providing qualitative insight into decision factors. For participants in the dashboard conditions (C and D), their configuration choices (data scope, permitted purpose, storage location, and retention period) were saved.

\paragraph{Covariates.} Demographic variables included age group, gender, preferred language, education level, and voting eligibility. These data were collected to describe the sample. A subset (age, gender, and education) served as covariates in the regression models.

\paragraph{Analysis strategy.} The data analysis followed a three-stage approach. First, chi-square tests were used to examine whether the donation rates differed across transparency levels, control levels, and the four conditions, with a Bonferroni correction (adjusted $\alpha = .017$). Second, binary logistic regression tested the three hypotheses through a sequence of nested models: Model~1 tested the transparency main effect (H1), Model~2 tested the control main effect (H2), Model~3 included both factors, Model~4 added the interaction term to test H3, and Model~5 included demographic covariates as a robustness check. Odds ratios (ORs) with 95\% confidence intervals (CIs) were used to quantify the effect sizes. Third, if the interaction term proved significant, a simple-effects analysis would examine the effect of transparency within each level of control, and vice versa. Effect sizes followed conventional thresholds: Cohen's $d$ values below 0.2 were considered negligible, 0.2--0.5 small, 0.5--0.8 medium, and above 0.8 large; for contingency tables, Cram\'er's $V$ values below 0.1 are negligible, 0.1--0.2 small, 0.2--0.4 medium, and above 0.4 large.

\paragraph{Technical validation.} To validate the experimental infrastructure and system architecture prior to human data collection, an automated testing framework based on the Random Silicon Sampling (RSS) methodology \citep{sun2024} was developed. Originally conceptualized as silicon sampling by \citet{argyle2023}, RSS enables the emulation of human population subgroups using group-level demographic distributions to verify that a platform's technical logic correctly handles diverse user profiles. The study simulated approximately 1{,}000 AI participants, a sample size that exceeded the minimum threshold of 200 synthetic respondents required for statistically stable opinion mirroring in silicon samples \citep{sun2024}. The RSS validation followed a diagnostic pipeline: (1)~extracting the demographic distribution of the target Swiss population, (2)~randomly assigning these attributes to configure 1{,}000 synthetic subjects, and (3)~prompting the Apertus LLM to simulate each persona and complete the full experimental journey. This approach verified the reliability of the condition assignment mechanisms, confirmed the correct contextual grounding of the information pipeline across multiple languages, and ensured that the PostgreSQL database remained stable with high concurrency. The complete measurement plan with all analysis phases is available from the corresponding author on reasonable request.

\section{Results}\label{sec:results}

\subsection{Sample and descriptive statistics}\label{subsec:descriptive}

A total of $N = 291$ participants started the online survey. Of these, 79 did not complete the post-task survey and were therefore excluded from the analysis, and an additional 7 participants failed the attention check and were excluded. No participants were missing condition assignment or donation decision data. This resulted in a final analytic sample of $N = 205$ and a completion rate of 70.4\% (Table~\ref{tab:flow}).

Randomization resulted in approximately balanced cell sizes, with minor variation due to the online recruitment process: Condition~A ($n = 50$, 24.4\%), Condition~B ($n = 53$, 25.9\%), Condition~C ($n = 45$, 22.0\%), and Condition~D ($n = 57$, 27.8\%). To verify that random assignment produced comparable groups, baseline measures collected before the experimental manipulation were compared across conditions using the Kruskal--Wallis test. No significant differences were found in terms of technology comfort ($H(3) = 0.59$, $p = .899$), privacy concern ($H(3) = 2.94$, $p = .401$), or ballot familiarity ($H(3) = 0.43$, $p = .934$), confirming successful randomization.

\begin{figure}[ht]
\centering
\includegraphics[width=0.95\textwidth]{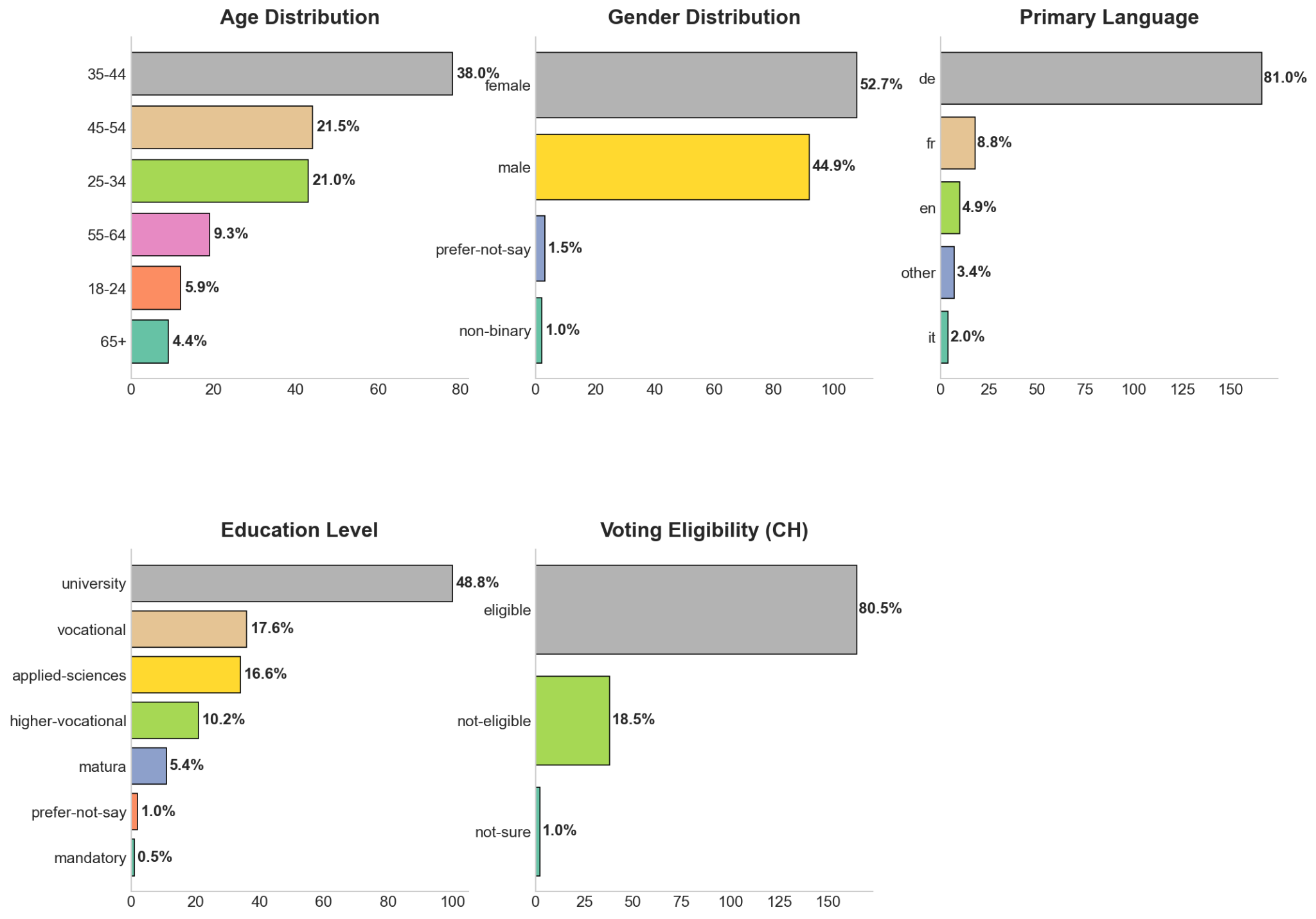}
\caption{Sample demographics: age, gender, primary language, education level, and voting eligibility ($N = 205$)}\label{fig:demo}
\end{figure}

The largest age group was 35--44 years (38.0\%), followed by 45--54 (21.5\%) and 25--34 (21.0\%) years. Female participants formed a slight majority (52.7\%), followed by male participants (44.9\%). German was the primary language for 81.0\% of the respondents, followed by French (8.8\%) and English (4.9\%). Nearly half of the respondents held a university degree (48.8\%), and 80.5\% were eligible to vote in Swiss elections (Figure~\ref{fig:demo}).

\paragraph{Donation rates.} The overall donation rate was 91.7\% (95\% CI [87.1, 94.8]). Donation rates were similarly high across all conditions, ranging from 90.0\% to 93.3\% (Table~\ref{tab:rates}); the narrow range of 3.3~percentage points between the conditions suggests minimal differentiation due to the experimental manipulations.

\begin{table}[ht]
\centering
\caption{Data donation rates by experimental condition}\label{tab:rates}
\begin{tabular}{@{}llcccc@{}}
\toprule
Condition & Description & $n$ & Donations & Rate (\%) & 95\% CI \\
\midrule
A & T0C0 (Baseline) & 50 & 45 & 90.0 & [78.6, 95.7] \\
B & T1C0 (Transparency only) & 53 & 48 & 90.6 & [79.7, 95.9] \\
C & T0C1 (Control only) & 45 & 42 & 93.3 & [82.1, 97.7] \\
D & T1C1 (Both) & 57 & 53 & 93.0 & [83.3, 97.2] \\
Overall & & 205 & 188 & 91.7 & [87.1, 94.8] \\
\bottomrule
\end{tabular}

\smallskip
{\footnotesize \textit{Note.} Confidence intervals use the Wilson score interval.}
\end{table}

\begin{figure}[ht]
\centering
\includegraphics[width=0.8\textwidth]{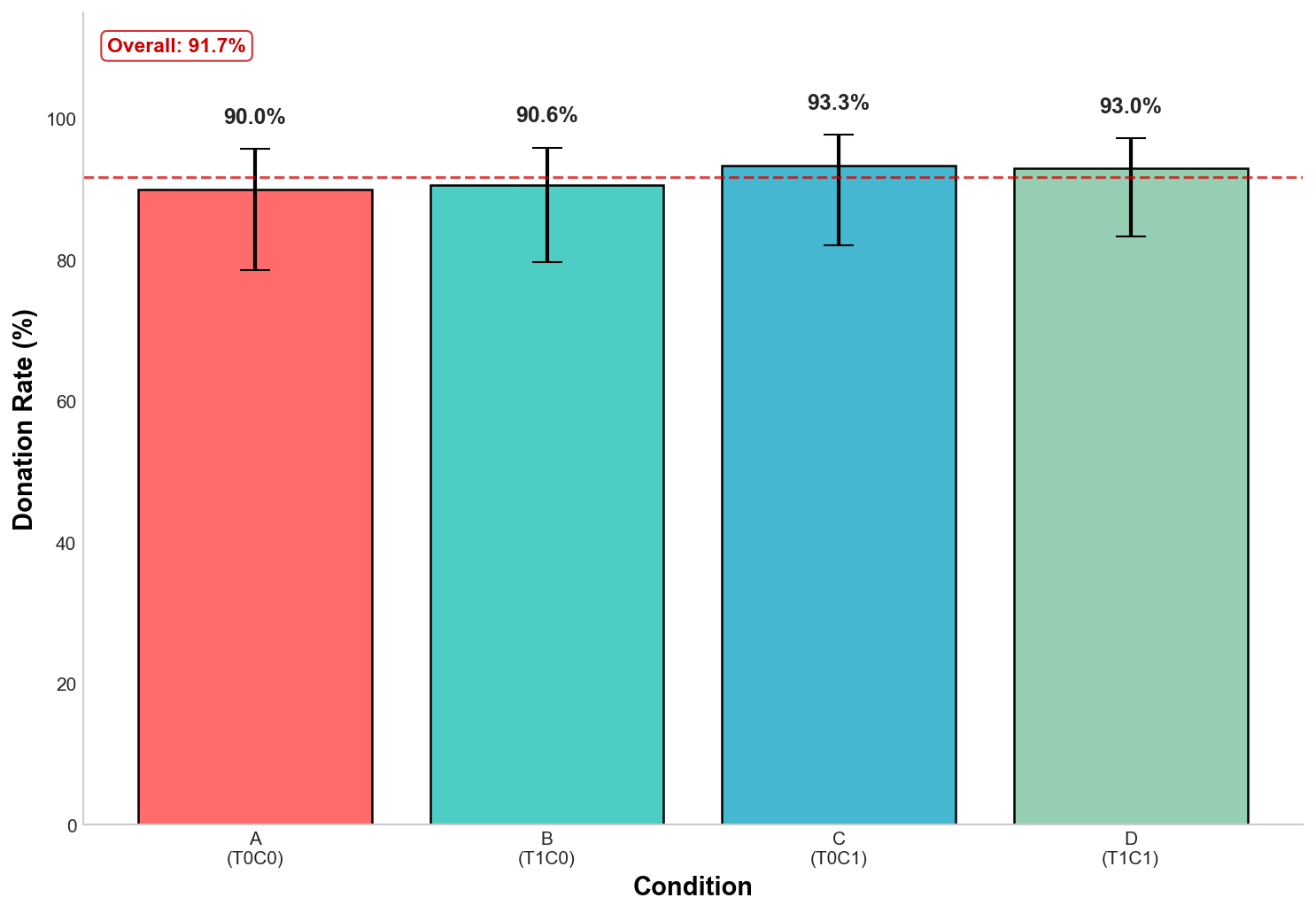}
\caption{Data donation rates by condition with 95\% confidence intervals. The dashed line marks the overall rate of 91.7\%}\label{fig:rates}
\end{figure}

When examined by factor level, the transparency effect was $+0.2$ percentage points (T1~$-$~T0), and the control effect was $+2.8$ percentage points (C1~$-$~C0). All confidence intervals overlapped substantially (Table~\ref{tab:factor}; Figure~\ref{fig:rates}).

\begin{table}[ht]
\centering
\caption{Donation rates by factor level}\label{tab:factor}
\begin{tabular}{@{}llcccc@{}}
\toprule
Factor & Level (conditions) & $n$ & Donations & Rate (\%) & 95\% CI \\
\midrule
Transparency & T0 low (A, C) & 95 & 87 & 91.6 & [84.3, 95.7] \\
             & T1 high (B, D) & 110 & 101 & 91.8 & [85.2, 95.6] \\
Control      & C0 low (A, B) & 103 & 93 & 90.3 & [83.1, 94.7] \\
             & C1 high (C, D) & 102 & 95 & 93.1 & [86.5, 96.7] \\
\bottomrule
\end{tabular}
\end{table}

The high donation rate was consistent across recruitment channels. The participants from the panel provider donated at 92.9\% and were predominantly female with vocational or applied-education backgrounds; they reported lower comfort with technology.

\paragraph{Risk perception and trust.} Risk perception (OUT-RISK) was moderate across conditions (overall $M = 3.12$, $SD = 1.46$ on a one-to-six scale), with condition~A showing the lowest risk perception ($M = 2.73$) and condition~C the highest ($M = 3.41$). Trust levels (OUT-TRUST) were relatively high and consistent (overall $M = 4.40$, $SD = 1.32$), with condition~D showing slightly higher trust ($M = 4.54$). Table~\ref{tab:risktrust} summarizes risk perception and trust by condition. Despite moderate risk perception, trust levels remained high across all conditions, suggesting that perceived risk did not substantially undermine confidence in the system.

\begin{table}[ht]
\centering
\caption{Risk perception and trust by condition}\label{tab:risktrust}
\begin{tabular}{@{}lcccc@{}}
\toprule
Condition & OUT-RISK $M$ & OUT-RISK $SD$ & OUT-TRUST $M$ & OUT-TRUST $SD$ \\
\midrule
A (T0C0) & 2.73 & 1.44 & 4.40 & 1.37 \\
B (T1C0) & 3.08 & 1.42 & 4.38 & 1.33 \\
C (T0C1) & 3.41 & 1.48 & 4.22 & 1.36 \\
D (T1C1) & 3.26 & 1.47 & 4.54 & 1.25 \\
Overall & 3.12 & 1.46 & 4.40 & 1.32 \\
\bottomrule
\end{tabular}
\end{table}

\subsection{Manipulation checks}\label{subsec:manip}

Manipulation checks were used to assess whether the experimental conditions produced the intended psychological states. Two manipulation checks were conducted: perceived transparency (MC-T) to validate the transparency manipulation (H1) and perceived control (MC-C) to validate the control manipulation (H2). Following the measurement plan, the Shapiro--Wilk test was used to assess normality; when normality assumptions were violated, Mann--Whitney $U$ tests were employed with rank-biserial $r$ as the effect size.

Transparency was assessed by comparing perceived transparency scores between low-transparency conditions (T0: A and C) and high-transparency conditions (T1: B and D). Shapiro--Wilk tests indicated non-normal distributions in both groups (T0: $W = 0.87$, $p < .001$; T1: $W = 0.87$, $p < .001$), so a Mann--Whitney $U$ test was conducted (Table~\ref{tab:mct}). The test indicated no significant difference in perceived transparency between conditions, $U = 5{,}518.0$, $p = .480$. Although participants in high-transparency conditions reported slightly higher perceived transparency ($M = 4.83$, $SD = 1.09$) than those in low-transparency conditions ($M = 4.67$, $SD = 1.24$), this difference was negligible (rank-biserial $r = 0.06$, 95\% CI [$-0.08$, 0.19]). The Data Nutrition Label did not significantly increase perceived transparency.

\begin{table}[ht]
\centering
\caption{Manipulation check for perceived transparency (MC-T)}\label{tab:mct}
\begin{tabular}{@{}lcccc@{}}
\toprule
Group & Conditions & $n$ & Mean & $SD$ \\
\midrule
T0 (low transparency) & A, C & 95 & 4.67 & 1.24 \\
T1 (high transparency) & B, D & 110 & 4.83 & 1.09 \\
\bottomrule
\end{tabular}
\end{table}

The Mann--Whitney $U$ test for control revealed a significant difference between conditions, $U = 6{,}325.0$, $p = .010$ (Table~\ref{tab:mcc}). Participants in high-control conditions reported significantly higher perceived control ($M = 4.84$, $SD = 0.95$) than those in low-control conditions ($M = 4.35$, $SD = 1.34$). The effect size was small (rank-biserial $r = 0.20$, 95\% CI [0.07, 0.33]), indicating that the granular consent dashboard effectively increased perceived control.

\begin{table}[ht]
\centering
\caption{Manipulation check for perceived control (MC-C)}\label{tab:mcc}
\begin{tabular}{@{}lcccc@{}}
\toprule
Group & Conditions & $n$ & Mean & $SD$ \\
\midrule
C0 (low control) & A, B & 103 & 4.35 & 1.34 \\
C1 (high control) & C, D & 102 & 4.84 & 0.95 \\
\bottomrule
\end{tabular}
\end{table}

\begin{figure}[ht]
\centering
\includegraphics[width=0.9\textwidth]{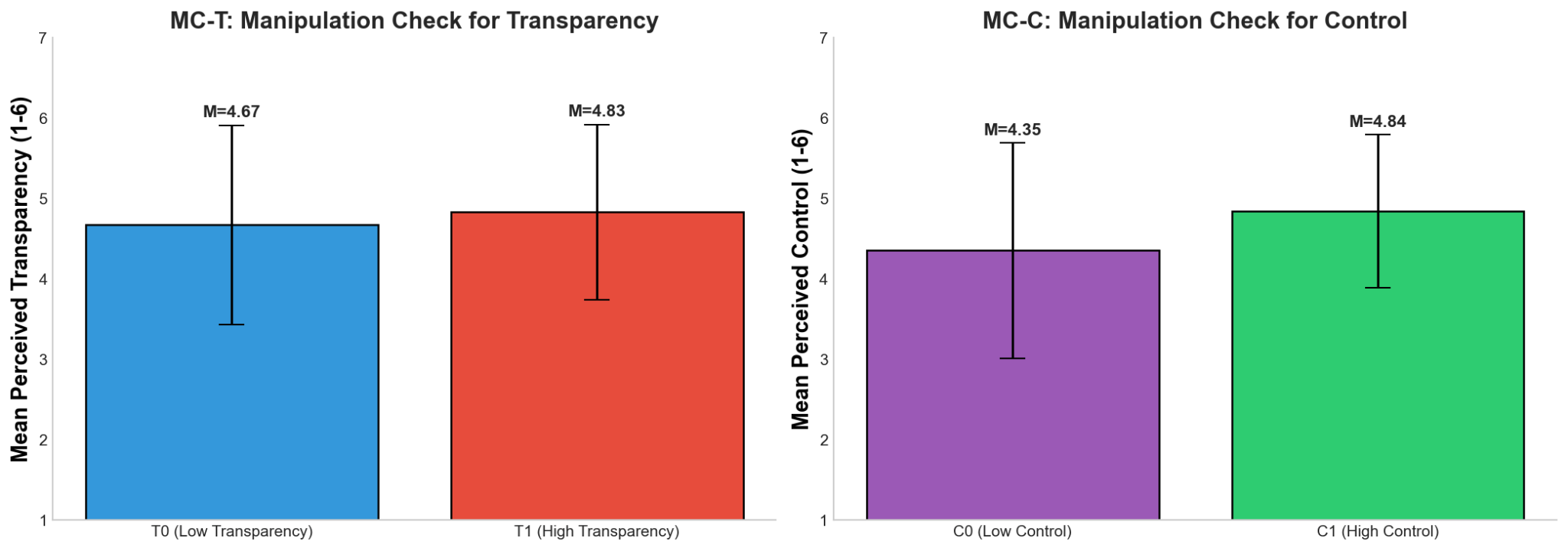}
\caption{Manipulation check results for perceived transparency and perceived control by factor level. Error bars represent one standard deviation}\label{fig:manip}
\end{figure}

The manipulation check for control (MC-C) was successful. However, the manipulation check for transparency (MC-T) did not reach statistical significance, suggesting that the Data Nutrition Label did not produce a detectable increase in perceived transparency. This partial manipulation failure has implications for interpreting the transparency hypothesis (H1) in the subsequent analyses (Table~\ref{tab:manip}; Figure~\ref{fig:manip}).

\begin{table}[ht]
\centering
\caption{Summary of manipulation check results}\label{tab:manip}
\begin{tabular}{@{}llcccl@{}}
\toprule
Manipulation & Comparison & $U$ & $p$ & $r$ & Result \\
\midrule
Transparency (MC-T) & T1 vs T0 & 5{,}518.0 & .480 & 0.06 & Not passed \\
Control (MC-C) & C1 vs C0 & 6{,}325.0 & .010 & 0.20 & Passed \\
\bottomrule
\end{tabular}

\smallskip
{\footnotesize \textit{Note.} $U$ is the Mann--Whitney statistic and $r$ is the rank-biserial effect size.}
\end{table}

\subsection{Hypothesis testing}\label{subsec:hyp}

To evaluate the three hypotheses, chi-square tests were conducted to analyze bivariate associations, followed by logistic regression models to estimate main and interaction effects while adjusting for relevant covariates. Chi-square tests of independence examined the relationship between the experimental conditions and data donation decisions. Three tests were performed using the Bonferroni-corrected significance threshold ($\alpha = .05/3 = .017$; Table~\ref{tab:chisq}).

\begin{table}[ht]
\centering
\caption{Chi-square tests: data donation by experimental condition}\label{tab:chisq}
\begin{tabular}{@{}llccccc@{}}
\toprule
Test & Comparison & $\chi^2$ & df & $p$ & Cram\'er's $V$ & Significant \\
\midrule
1 & Transparency (T0 vs T1) & 0.00 & 1 & 1.000 & 0.000 & No \\
2 & Control (C0 vs C1) & 0.24 & 1 & .627 & 0.034 & No \\
3 & Condition (A--D) & 0.56 & 3 & .905 & 0.052 & No \\
\bottomrule
\end{tabular}
\end{table}

None of the three chi-square tests reached statistical significance. The transparency effect (Test~1) showed virtually no association, $\chi^2(1) = 0.00$, $p = 1.000$, $V = 0.000$. The control effect (Test~2) was also non-significant, $\chi^2(1) = 0.24$, $p = .627$, $V = 0.034$. The omnibus condition effect (Test~3) was non-significant, $\chi^2(3) = 0.56$, $p = .905$, $V = 0.052$. All effect sizes were negligible, consistent with the high baseline rate reported in Section~\ref{subsec:descriptive}.

To test all three hypotheses at once and control for potential confounding factors, a series of nested logistic regression models were estimated: Model~1 (transparency only), Model~2 (control only), Model~3 (main effects: T~+~C), Model~4 (interaction: T~+~C~+~T$\times$C), and Model~5 (full model with covariates). Table~\ref{tab:models} compares the model fit across all five specifications.

\begin{table}[ht]
\centering
\caption{Logistic regression model comparison}\label{tab:models}
\begin{tabular}{@{}lcccccc@{}}
\toprule
Model & Parameters & Log-likelihood & AIC & LRT $\chi^2$ & df & $p$ \\
\midrule
M1 (T only) & 2 & $-58.60$ & 121.2 & -- & -- & -- \\
M2 (C only) & 2 & $-58.33$ & 120.7 & -- & -- & -- \\
M3 (T + C) & 3 & $-58.33$ & 122.7 & 0.55 & 1 & .460 \\
M4 (T + C + T$\times$C) & 4 & $-58.32$ & 124.6 & 0.01 & 1 & .908 \\
M5 (Full) & 7 & $-51.91$ & 117.8 & 12.82 & 3 & .005 \\
\bottomrule
\end{tabular}
\end{table}

Model comparison reveals that neither adding control to transparency (M3 vs M1, $\chi^2(1) = 0.55$, $p = .460$) nor the interaction term (M4 vs M3, $\chi^2(1) = 0.01$, $p = .908$) significantly improves model fit. However, adding demographic covariates (M5 vs M4) significantly improved the fit, $\chi^2(3) = 12.82$, $p = .005$, suggesting that demographic variables explain more variance in donation behavior than the experimental manipulations. Based on the AIC, Model~5 provided the best fit (AIC~=~117.8). Table~\ref{tab:coef} reports the coefficients for the main-effects model (M3).

\begin{table}[ht]
\centering
\caption{Logistic regression coefficients: main-effects model (M3)}\label{tab:coef}
\begin{tabular}{@{}lccccc@{}}
\toprule
Predictor & $\beta$ & SE & OR & 95\% CI & $p$ \\
\midrule
Intercept & 2.22 & 0.42 & 9.23 & -- & -- \\
Transparency (T1) & 0.01 & 0.51 & 1.02 & [0.38, 2.75] & .977 \\
Control (C1) & 0.38 & 0.51 & 1.46 & [0.53, 4.00] & .463 \\
\bottomrule
\end{tabular}

\smallskip
{\footnotesize \textit{Note.} Reference categories are T0 for transparency and C0 for control.}
\end{table}

The full model (M5), including demographic covariates, significantly improved the fit over Model~4 (Table~\ref{tab:models}; LRT $\chi^2 = 12.82$, $p = .005$). Age was a significant predictor ($\mathrm{OR} = 0.52$, $p = .004$), indicating that older participants were less likely to donate than younger participants. Education was also significant ($\mathrm{OR} = 1.53$, $p = .031$), with each higher education level associated with 1.53 times the odds of donation. The experimental variables, transparency and control, remained non-significant predictors in the full model.

In sum, H1 (transparency effect) was not supported: transparency did not significantly predict donation, $\beta = 0.01$, $\mathrm{OR} = 1.02$, 95\% CI [0.38, 2.75], $p = .977$; the odds ratio is essentially 1.0, indicating no effect of the Data Nutrition Label on donation behavior. H2 (control effect) was not supported: control did not significantly predict donation, $\beta = 0.38$, $\mathrm{OR} = 1.46$, 95\% CI [0.53, 4.00], $p = .463$; although the direction favors higher donation in the high-control condition, the wide confidence interval spanning 1.0 precludes meaningful interpretation. H3 (interaction effect) was not supported: the interaction term was not significant in M4, $\mathrm{OR} = 0.89$, 95\% CI [0.12, 6.73], $p = .908$; no synergistic or antagonistic interaction between transparency and control was detected (Table~\ref{tab:hyp}).

\begin{table}[ht]
\centering
\caption{Summary of hypothesis tests and effect sizes}\label{tab:hyp}
\begin{tabular}{@{}llcccl@{}}
\toprule
Hypothesis & Effect & OR & 95\% CI & $p$ & Decision \\
\midrule
H1 & Transparency & 1.02 & [0.38, 2.75] & .977 & Not supported \\
H2 & Control & 1.46 & [0.53, 4.00] & .463 & Not supported \\
H3 & T $\times$ C interaction & 0.89 & [0.12, 6.73] & .908 & Not supported \\
\bottomrule
\end{tabular}
\end{table}

None of the three hypotheses were supported. All treatment effects were negligible in magnitude, consistent with the non-significant $p$-values. Although the control manipulation successfully increased perceived control (Section~\ref{subsec:manip}), it did not translate into measurable differences in data donation behavior.

\paragraph{Bayesian robustness check.} To distinguish evidence of absence from absence of evidence, Bayes factors were computed for each hypothesis using a beta-binomial conjugate model with uninformative Beta(1, 1) priors. For transparency, the data provided strong evidence for the null ($\mathrm{BF}_{01} = 10.30$), moderate evidence for control ($\mathrm{BF}_{01} = 7.99$), and extreme evidence for the four-condition comparison ($\mathrm{BF}_{01} = 428.76$). Following the classification of \citet{jeffreys1961}, these results confirm that the non-significant frequentist tests reflect the genuine absence of treatment effects on donation behavior rather than insufficient statistical power.

\subsection{Exploratory analysis}\label{subsec:explore}

Participants in the high-control conditions (C and D, $n = 102$) configured their data donations using the granular consent dashboard. Chi-square tests revealed no significant differences between conditions C and D for any dashboard dimension (all $p > .38$; all $V < .14$). The most common configurations were full data scope (73.5\%), academic-only purpose (67.6\%), Swiss server storage (47.1\%), and retention until purpose fulfilled (59.8\%). Notably, 68.6\% preferred Swiss or EU server storage, suggesting concerns regarding data sovereignty (Table~\ref{tab:config}). Even though most participants donated, they clearly constrained the terms of use.

\begin{table}[ht]
\centering
\caption{Most common dashboard configuration choices in the high-control conditions ($n = 102$)}\label{tab:config}
\begin{tabular}{@{}llc@{}}
\toprule
Configuration dimension & Most frequent choice & Share (\%) \\
\midrule
Data scope & Full conversations & 73.5 \\
Permitted purpose & Academic research only & 67.6 \\
Storage location & Swiss servers only & 47.1 \\
Storage location & Swiss or EU servers & 68.6 \\
Retention period & Until research purpose fulfilled & 59.8 \\
\bottomrule
\end{tabular}

\smallskip
{\footnotesize \textit{Note.} Shares of participants in conditions C and D who configured the dashboard. The two storage rows are not mutually exclusive: 47.1\% chose Swiss-only, while 68.6\% accepted Swiss or EU storage.}
\end{table}

\begin{figure}[ht]
\centering
\includegraphics[width=0.9\textwidth]{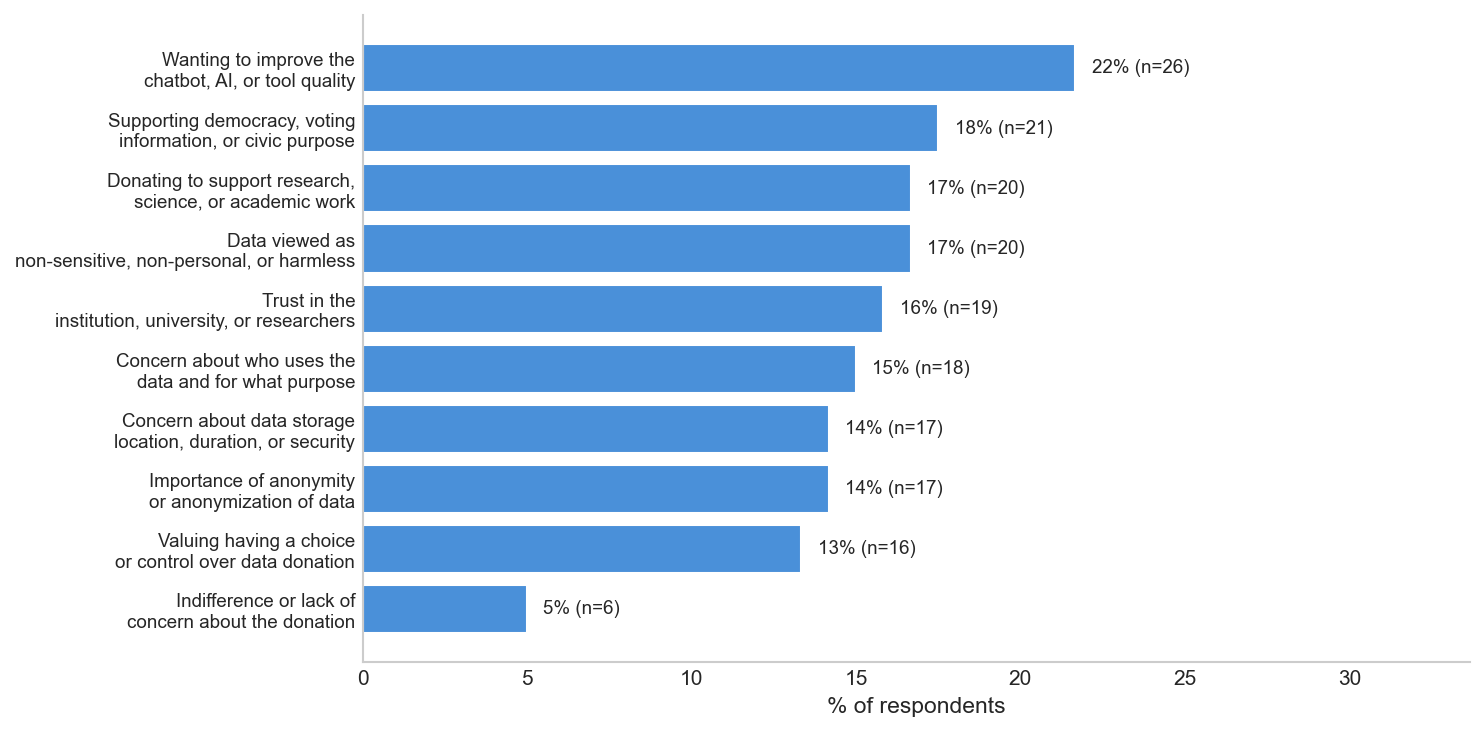}
\caption{Frequencies of the open-ended response themes ($N = 120$ respondents), based on deterministic multilingual keyword matching}\label{fig:themes}
\end{figure}

\paragraph{Open-ended feedback.} The 120 open-ended responses to the question ``What mattered most for your data donation decision?'' were analyzed using a deterministic multilingual keyword-based content analysis. Ten themes were derived inductively from a reading of all responses. For each theme, keyword lists were compiled in German, French, and English, the three primary participant languages. Coding was performed through case-insensitive substring matching with multi-label assignment (responses could match multiple themes), which ensured full reproducibility. Because the coding was algorithmic rather than manual, traditional inter-rater reliability metrics do not apply; the codebook and all response-level codings are provided as reproducibility artifacts so that the results can be verified. The codebook achieved 93.3\% coverage (112/120 responses matched at least one theme; $M = 1.50$ themes per response). Figure~\ref{fig:themes} shows the theme frequencies, and Table~\ref{tab:quotes} shows representative quotes for each theme. The most frequently mentioned themes were improving AI or chatbot quality (21.7\%), supporting democracy or a civic purpose (17.5\%), supporting research (16.7\%), and perceiving the data as non-personal (16.7\%).

\begin{table}[ht]
\centering
\caption{Representative quotes by theme}\label{tab:quotes}
\begin{tabular}{@{}llp{0.58\textwidth}@{}}
\toprule
Theme & Cond. & Quote \\
\midrule
Improve AI & D & ``I think it is important to help research of open source models'' \\
Civic purpose & D & ``Unterst\"utze die Informationsbeschaffung zu Abstimmungen'' \\
Support research & B & ``Datenspende hilft der Forschung und am Ende uns allen'' \\
Data not personal & C & ``The chat did not contain information that I consider to be of personal nature'' \\
Trust in institution & B & ``That it is handled by institutions that I trust'' \\
Control/choice & A & ``Mir wurde eine einfache Wahlm\"oglichkeit gegeben'' \\
\bottomrule
\end{tabular}

\smallskip
{\footnotesize \textit{Note.} One representative response per theme. Original languages are preserved.}
\end{table}

Participants in the dashboard-only condition (C) mentioned control and choice as decision factors far more frequently (37.0\%) than in any other condition (0.0\%--10.3\%). When the Data Nutrition Label was added alongside the dashboard (condition~D vs C), mentions of control dropped from 37.0\% to 10.0\%, whereas references to supporting research and data purpose increased substantially. This suggests that the DNL redirected attention from the act of choosing to the substance of what was being disclosed. These patterns were exploratory: the subgroup sizes ($n = 27$--$34$ per condition) precluded statistical testing, and the differences should be interpreted as hypothesis-generating for future research rather than as evidence of treatment effects.

\section{Discussion}\label{sec:discussion}

\subsection{Interpretation of results}\label{subsec:interpret}

The primary objective of this study was to investigate the causal influence of transparency and user control on the willingness of Swiss residents to donate their political chatbot data. Contrary to the proposed hypotheses, the experimental manipulations did not yield statistically significant differences in donation behavior.

Willingness to donate was uniformly high across all conditions. The baseline condition without enhanced transparency or control yielded a donation rate of 90.0\%, whereas the overall donation rate across the sample was 91.7\%. As described in Section~\ref{subsec:descriptive}, the donation rate did not have sufficient variance to detect treatment differences. This suggests that participants entered the decision situation with a pre-existing disposition to contribute data, driven by contextual factors rather than interface design. This study was presented as a Swiss academic project with a civic purpose and no commercial interest. That framing probably raised perceived legitimacy and public benefit enough that small differences in interface design did not matter much to participants.

The open-ended responses point in the same direction as the quantitative results. Many participants wrote that they wanted to support research, improve AI, or contribute to democratic processes. They saw donating as something positive, not as a risk to manage. The Privacy Paradox was developed mostly in commercial settings and may not transfer well to a context where people feel they are contributing to a public good.

On the psychological side, the manipulation checks add further detail. The granular consent dashboard made participants feel more in control, which means that the manipulation worked as intended. When people are already willing to donate, giving them additional control over the terms does not seem to push them further.

The Data Nutrition Label, on the other hand, did not raise perceived transparency. Scores were already high in the conditions without the label, which suggests that participants felt sufficiently informed by the baseline text alone. The label did not add enough to make a measurable difference. This does not reject transparency as a principle, but it does mean that the label offered little additional informational value in this particular setup.

Beyond donation decisions, the dashboard configuration data provide insights into user preferences when control is offered. Participants frequently restricted data use to academic purposes, preferred Swiss or European data storage, and selected finite retention periods. These choices indicate that, despite their high willingness to donate, participants exercised restraint when specifying the terms of use. High trust in the system does not automatically translate into unconditional consent. People still care about having control over how their data are shared, even if that control does not seem to change whether they share their data at all. This distinction is important.

Finally, it needs to be acknowledged that the donation was not real. Participants were not aware of this during the task, but no data were stored. This study was conducted in a controlled research setting; outside of it, the stakes are higher. Political queries stored somewhere and fed into training data over time are a different situation entirely. Whether transparency and control mechanisms would have the same effect under these conditions is an open question.

\subsection{Limitations}\label{subsec:limits}

The following limitations affect the interpretation and generalizability of the study's results. First, the ceiling effect (Section~\ref{subsec:descriptive}) constrained the outcome variability and statistical power. While this reflects a strong willingness to donate, it constrains causal inference about the effectiveness of the interface mechanisms tested.

Second, the institutional setting in which the study was embedded may have affected the way participants evaluated trust. Although the study was conducted by a student from a German university, it was framed as a non-commercial civic AI initiative dedicated to Swiss democratic participation, including hosting on sovereign Swiss infrastructure and using the Apertus language model developed by ETH Zurich and EPFL. These elements may have contributed to a high baseline trust that does not generalize to other institutional or international contexts.

Third, the transparency manipulation did not significantly increase perceived transparency. While the control manipulation functioned as intended, the absence of a psychological effect of transparency limits the internal validity of the first hypothesis. Therefore, the non-significant behavioral effect in this condition must be interpreted with caution.

Fourth, the study relied on a binary behavioral outcome, whether participants donated or not, which may have obscured more granular differences in user attitudes. Although the consent dashboard data provide some nuance, the main outcome variable lacks sensitivity to partial willingness, hesitation, or conditional preferences. For example, a Likert-based willingness scale could have captured varying degrees of hesitation that the binary measure collapsed into a single category.

Fifth, the civic framing of this study may have created a social desirability bias. The study was presented as academic research supporting Swiss democracy, which makes it socially difficult to refuse the donation. This acts as a demand characteristic, separate from the trust factors in Section~\ref{subsec:interpret}. The high donation rate under all conditions supports this reading. Participants did not know that the donation was simulated; therefore, the social pressure to comply was real for them.

\subsection{Future research}\label{subsec:future}

The high baseline trust observed in this study is probably connected to the academic and civic framing as well as the focus on national infrastructure. It would be worth researching how transparency and control mechanisms would work in contexts where institutional trust is low. The results may differ considerably in commercial, international, or politically polarized environments.

Future research could also explore the dynamic implementation of transparency and control. In this study, both the Data Nutrition Label and the granular consent dashboard were static, and participants saw them only once at the time of the donation decision. An alternative approach would be to directly link transparency to the actual behavior of the system. For example, a performance module could track how often the model's outputs match verified facts and display this accuracy to participants over time. In this way, trust could grow from what the system actually does, not from a single label shown once.

Along similar lines, longitudinal study designs would help capture how donation behavior develops over repeated interactions. Because data donation depends on continuous user engagement, tracking behavior over sequential multi-turn interactions would clarify how performance transparency shapes trust development, fatigue, and willingness to donate over time. Frameworks for structuring human-AI collaboration in civic engagement processes \citep{overney2025} could inform how transparency and control mechanisms are integrated into sustained participatory systems.

This study used a binary yes/no donation decision, which worked for hypothesis testing but left open how people actually felt about donating. Future studies should look at graded willingness or at whether people change their minds and revoke consent afterward. Helpful insights could also come from additional qualitative research on the factors that guide behavior; for that, a mixture of methods such as structured interviews or reflective in-the-moment questions could be employed. This would lead to a better understanding of how an individual processes framing and reaches a decision.

Finally, the premise of this study, that civic AI requires large amounts of donated data, may change. Approaches like synthetic reasoning \citep{abdin2024} and permissive-first provenance \citep{nguyen2025} use smaller amounts of human data as seeds rather than storing raw conversations. It could be valuable to find out whether residents are more willing to contribute when they know that their data are only used as seeds for synthetic pipelines and not stored in raw form. The focus would then shift from collecting data at scale to actively involving residents in shaping how the models are trained.

\section{Conclusion}\label{sec:conclusion}

This study investigated whether transparency and user control increase the willingness of Swiss residents to donate sensitive chatbot data for open-source AI development. In a $2\times2$ between-subjects experiment, 205 participants interacted with a chatbot built on the Apertus language model, and it was tested whether a visual Data Nutrition Label and a granular consent dashboard would encourage them to donate data in a democratic society. Neither a Data Nutrition Label nor a granular consent dashboard changed donation behavior: rates were high across all conditions (91.7\% overall), and Bayesian robustness checks confirmed genuine null effects. The dashboard did increase perceived control, and participants used it to restrict purpose, storage, and retention, so control shaped the terms of sharing rather than the decision to share. The most parsimonious explanation for this near-universal willingness is that both sides of the privacy calculus pointed in the same direction: a high perceived public benefit coincided with a low perceived risk, because many residents did not regard their anonymized queries as personal.

For practitioners who design public civic AI systems, these results indicate that who asks, for what purpose, and under what governance may matter more than the consent interface itself. Transparency and control remain important design principles, but their value may lie less in increasing participation rates and more in enabling residents to define the terms of their contribution. The dashboard configuration data clearly show that even when willingness to donate is near universal, residents actively restrict scope, purpose, and retention. Designing informed consent for civic AI therefore means designing for negotiation.

\bibliographystyle{plainnat}
\bibliography{references}

\end{document}